\documentclass[hyper]{JHEP3}

\input{epsf}
\usepackage{epsfig}
\usepackage{amssymb}
\usepackage{amsfonts}
\usepackage{amsbsy}

\def\IC{\mathbb{C}}

\def\IP{\mathbb{P}}

\def\CN {{\cal N}}

\def\CD {{\cal D}}

\def\CO {{\cal O}}

\def\la{\langle}
\def\ra{\rangle}
\def\half{\frac{1}{2}}

\newcommand{\eq}[1]{Eq.~(\ref{eq:#1})}

\renewcommand{\Im}{{\rm Im }}

\def\one{{\hbox{ 1\kern-.8mm l}}}

\def\p{\partial}
\def\ba{\bar{a}}
\def\bb{\bar{b}}
\def\bc{\bar{c}}
\def\bd{\bar{d}}

\def\bp{\bar{p}}
\def\bz{\bar{z}}

\def\bA{\bar{A}}

\def\bpartial{\bar{\partial}}

\def\bpsi{\bar{\psi}}

\def\bs{\bar{s}}

\def\bx{\bar{x}}

\def\bz{\bar{z}}

\def\bepsilon{\bar{\epsilon}}
\def\bA{\bar{A}}

\def\bra#1{{\langle}#1|}
\def\ket#1{|#1\rangle}

\def\vev#1{\langle{#1}\rangle}

\def\p{\partial}

\def\del{\Delta}
\def\ddel{{}^\bullet\! \Delta}
\def\deld{\Delta^{\hskip -.5mm \bullet}}
\def\dddel{{}^{\bullet \bullet} \! \Delta}
\def\ddeld{{}^{\bullet}\! \Delta^{\hskip -.5mm \bullet}}

\def\idt{\int \!\! d\tau\ }

\def\idtds{\int \!\! \int \!\! d\tau d\sigma\ }

\def\gam{\Gamma}

\def\idt{\int \!\! d\tau\ }

\def\idtds{\int \!\! \int \!\! d\tau d\sigma\ }

\preprint{RUNHETC-2008-15\\ 
ITEP-TH-37/08}

\title{Bergman Kernel from Path Integral}

\author{Michael R. Douglas$^{1,2,3}$ and Semyon Klevtsov$^{1,2,4}$\\
\\
$^1$  Simons Center for Geometry and Physics, 
Stony Brook University,\\
Stony Brook, NY 11794--3840, USA\\
\\
$^2$ NHETC and Department of Physics and Astronomy,
Rutgers University,\\
Piscataway, NJ 08855--0849, USA\\
\\
$^3$ I.H.E.S., Le Bois-Marie, Bures-sur-Yvette, 91440, France\\
\\
$^4$ ITEP, Moscow, 117259, Russia\\
\\
{\tt mrd@physics.rutgers.edu, klevtsov@physics.rutgers.edu} }

\abstract{
We rederive the expansion of the Bergman kernel on K\"ahler manifolds
developed by Tian, Yau, Zelditch, Lu and Catlin, using path integral and
perturbation theory, and generalize it to supersymmetric quantum mechanics.

One physics interpretation of this result is as an expansion of the
projector of wave functions on the lowest Landau level, in the special
case that the magnetic field is proportional to the K\"ahler form.
This is relevant for the quantum Hall effect in curved space,
and for its higher dimensional generalizations.
Other applications include the theory of coherent states, the study of
balanced metrics, noncommutative field theory, and a conjecture on
metrics in black hole backgrounds discussed in \cite{DK}.
We give a short overview of these various topics.

From a conceptual point of view, this expansion is noteworthy as it is
a geometric expansion, somewhat similar to the DeWitt-Seeley-Gilkey {\it et al}
short time expansion for the heat kernel, but in this case describing
the long time limit, without depending on supersymmetry.}

\begin{document}

\section{Introduction}

A prototypical topic at the interface of geometry and theoretical
physics is the study of quantum mechanics in curved space, {\it i.e.}
on a Riemannian manifold $M$ \cite{DeWitt:1965jb,Gilkey:1995mj,Kleinert:2004ev,Bastianelli:2006rx}. Many results in this area
are of great interest both to physicists and to mathematicians, with
some examples being the DeWitt-Seeley-Gilkey short time expansion of
the heat kernel, and the relation between supersymmetric quantum
mechanics and the Atiyah-Singer index theorem \cite{AlvarezGaume:1983ig,AlvarezGaume:1983at,Friedan:1983xr}.

A more recent result which, although not well known by physicists,
we feel also belongs in this category, is the expansion for the
Bergman kernel on a K\"ahler manifold developed by
Tian, Yau, Zelditch, Lu and Catlin \cite{Tian:metrics,Zelditch:Szego,Catlin,Lu}.  
It applies to K\"ahler quantization and gives an asymptotic expansion around the
semiclassical limit.  This has many uses in mathematics
\cite{Tian:metrics,Donaldson1,Donaldson2,Donaldson:numeric};
see the recent book \cite{MaMarinescu}.

Here we will provide a physics derivation of the asymptotic expansion of the Bergman kernel using path integrals, and explain various possible applications of this result.

In physics terms, perhaps the simplest way to define the Bergman
kernel is in the context of quantum mechanics of a particle in a
magnetic field, in which it is the projector on the lowest Landau
level.  It is not hard to see that the limit of large magnetic field
is semiclassical, so that one can get an expansion in the inverse
magnetic field strength using standard perturbative methods.

Our basic result is to rederive the Tian-Yau-Zelditch {\it et al} expansion as the
large time limit of the perturbative expansion for the quantum
mechanical path integral.  We also generalize it to $N=1$ and $N=2$
supersymmetric quantum mechanics.

Let us state the basic result for (nonsupersymmetric) quantum mechanics.
We consider a compact K\"ahler manifold $M$, and a particle in magnetic field,
with the field strength proportional to the K\"ahler form on the manifold
\begin{equation}
\label{Fomega}
F_{ij} \sim \omega_{ij} .
\end{equation}
One can show (see below) that, just as for a constant magnetic field in
flat space, in this situation the spectrum is highly degenerate,
splitting into ``Landau levels.''  Let the 
lowest Landau level (LLL or ground state) be $N$-fold degenerate with a basis
of orthonormal wave functions $\psi_I(x)$, then
we define the projector on the LLL as
\begin{equation}
\label{eq:def-rho}
\rho(x,x') = \sum_{I=1}^N \psi^*_I(x') \psi_I(x) .
\end{equation}
We could also regard this as a density matrix describing a mixed state
in which each ground state appears with equal weight, describing the
the zero temperature state of maximum entropy.

We then consider scaling up the magnetic field by a parameter $k$,
as $F \rightarrow k F$.  Note that on a compact
manifold, $F$ must satisfy a Dirac quantization condition; thus we 
take $k=1,2,3,\ldots$
In the large $k$ limit, the diagonal term then satisfies
\begin{equation}
\label{exp1}
\rho(x,x) \sim
 k^n\left(1+\frac{\hbar}{2k}R+\frac{\hbar^2}{k^2}\left(\frac13\Delta
 R+\frac1{24}|\mbox{Riem}|^2-\frac16|\mbox{Ric}|^2+\frac18R^2\right)
 +\CO((\hbar/k)^3)\right)
\end{equation}
as an asymptotic expansion \cite{Lu} (see Appendix A for the precise definitions of different terms here).

In some ways this expansion is similar to the well known short time
expansion of the heat kernel, but note that it is a long time
expansion, because it projects on the ground states.  Unlike other
analogous results for ground states, it does not require supersymmetry,
either for its definition or computation.  Of course, similar results
can be obtained for supersymmetric theories, our point is that that they
do not depend on supersymmetry (whether they depend ultimately on
holomorphy is an interesting question discussed below).

There are various other physics interpretations of this result.  One
familiar variation is to regard $M$ as a phase space, and try to
quantize it, following Berezin \cite{Berezin}.  As a phase space, $M$
must have a structure which can be used to define Poisson brackets; it
is familiar \cite{Woodhouse} that this is a symplectic structure, {\it
i.e.} a nondegenerate closed two-form $\omega$.  The definitions we
just gave are then the standard recipe of geometric quantization
\cite{Rawnsley}.  They lead to a finite dimensional Hilbert space,
whose dimension is roughly the phase space volume of $M$ in
units of $(2\pi\hbar)^n$.  In this interpretation, the parameter $k$ plays
the role of $1/\hbar$, and thus the large $k$ limit is semiclassical.

From this point of view, it is intuitive that one should be able to
localize a wave function in a region of volume $(2\pi\hbar)^n \sim
1/k^n$, and thus in the large $k$ limit
the density matrix $\rho(x,x)$ should be computable in
terms of the local geometry and magnetic field near $x$.

To do this, given a point $z \in M$, one might seek a wave function
$\psi_z(z')$ which is peaked around $z$.
Given an orthonormal basis for $\cal H$, a natural candidate is
$$
\psi_z(z') = \sum_I \psi_I^*(\bar z) \psi_I(z')
$$
This is a coherent state (in the sense of \cite{Rawnsley}).  It can be
used to define the symbol of an operator, the star product
\cite{Reshetikhin}, and related constructions.
In this context the Bergman kernel is the
``reproducing kernel'' studied in \cite{Klauder:1989qp}, see
\cite{Woodhouse} for a review. For recent work on applications of the
Bergman kernel to quantization of K\"ahler manifolds see
\cite{Kirwin,Lukic:2007nc}.  Another
recent paper discussing the topic is \cite{IuliuLazaroiu:2008pk}.

Our original interest in this type of result came from the study of
balanced metrics in \cite{Donaldson:numeric}, and a conjecture about their relevance
for black holes in string theory stated in \cite{DK}.  However, after
realizing that these results and techniques do not seem to have direct
analogs in the physics literature and could have other applications,
we decided to provide a more general introduction as well.

\section{Background}

Let us give a few mathematical
and physical origins and applications of this type of result.

\subsection{Particle in a magnetic field}

We consider a particle of mass
$m$ (which later we set to one) and charge $k$, moving on a $2n$-dimensional manifold $M$ which
carries a general metric $g_{ij}$, and a magnetic field $F_{ij}$.
It is described by a wave function $\psi(x;t)$ which satisfies
the Schr\"odinger equation,
\begin{equation} \label{eq:schro-one}
H\psi \equiv
 \frac{\hbar^2}{2 m\sqrt{g}} D_i ~\sqrt{g} g^{ij}~ D_j \psi = E\psi ,
\end{equation}
where $D_i=i\partial_i + kA_i$ is the covariant derivative appropriate
for a scalar wavefunction with charge $k$, and $E=i\hbar\partial/\partial t$.
If $M$ is topologically nontrivial, as usual
we need to work in coordinate patches
related by gauge and coordinate transformations, and this expression 
applies in each coordinate patch.
We can of course also
consider the time-independent Schr\"odinger equation with $E$ fixed,
and seek the energy eigenstates $H\psi_i(x)=E_i\psi_i$.

Let us now consider the limit of large magnetic field or equivalently
large $k$.  The case of two-dimensional
Euclidean space $g_{ij}=\delta_{ij}$ with a
constant magnetic field $F_{ij}=B \epsilon_{ij}$
is very familiar.  The energy eigenstates
break up into Landau levels, such that all states in the $l$'th level have
energy $E_l=\hbar k B(l+\half)/m$.  Within a Landau level, one can
roughly localize a state to a region of volume $\hbar/k B$.

These results can be easily generalized to $d=2n$ dimensions.
Choose coordinates such that the magnetic field lies in the $12$, $34$
planes and so forth, and $B_{12}>0$, $B_{34}>0$ etc.
Then, considering the lowest Landau level (LLL) we have 
\begin{equation} \label{eq:LLL-energy}
E=\frac{\hbar}{2m} (B_{12}+\ldots+B_{2n-1,2n}),
\end{equation}
with states localized as before within each two-plane.

In a general metric and magnetic field, while one might not at first
expect this high degree of degeneracy, it is still possible.
When the magnetic field is much larger than the curvature of the
metric, the intuition that wave functions localize should still be
valid.  Then, we might estimate the energy of a wave function in the
lowest Landau level localized around a point $x$ as \eq{LLL-energy},
where the components $B_{12}$, $B_{34}$ and so on are evaluated in a
local orthonormal frame.  If the energy $E$ in 
\eq{LLL-energy} is constant, then all states
in the LLL will be degenerate, at least in the limit of large $k$.

The proper generalization of the splitting of the components of $B$ \eq{LLL-energy}
for nonconstant magnetic fieds seems to be, that the field strength should be nonzero only for mixed components $F_{a\ba}$, with $F_{ab}=F_{\ba\bb}=0$ in the complex coordinates $z^a,\,\bz^{\ba}$ ($a,\ba=1,...,n$) on the manifold. Mathematically it means, that the underlying line bundle is holomorphic. In this case,
the argument can be sharpened by using the identity
$$
[D_i,D_j] = F_{ij} .
$$
to rewrite the Hamiltonian as
$$
H = g^{a\ba}F_{a\ba} + g^{a\ba} D_{a} {\bar D}_{\ba}.
$$
This makes it clear that if the following combination is constant
\begin{equation}
\label{const}
g^{a\ba}F_{a\ba}={\rm const}
\end{equation}
every wave function satisfying
\begin{equation} \label{eq:Dbar-eqns}
0 = {\bar D}_{\ba} \psi
\end{equation}
will be degenerate and lie in the LLL.  This argument can hold away
from the strict $k\rightarrow\infty$ limit. 

The condition (\ref{const}) is known as hermitean Yang-Mills equation, and is essentially equivalent to Maxwell equation in the case $F^{0,2}=0$. Recalling, that the metric coefficients on the K\"ahler manifold
are related to the K\"ahler form $\omega$ as $g_{a\ba}=-i\omega_{a\ba},\,g_{\ba a}=i\omega_{\ba a}$, one can see that our choice (\ref{Fomega}) of the magnetic field strength
\begin{equation}
\label{eq:CS}
F_{a\ba}=k g_{a\ba}
\end{equation}
does satisfy the condition (\ref{const}). 

In fact the previous argument relies only upon the Maxwell equations and the condition $F^{2,0}=0$. This suggests that there exist more general magnetic field configurations, than (\ref{eq:CS}), for which the
LLL is still highly degenerate and the expansion in large magnetic fields, analogous to (\ref{exp1}), is possible. For example this includes the case when $b^{1,1}(M)>1$. We will elaborate this question in the future publication.

The previous physical condition for the field strength is equivalent to the mathematical
condition that $M$ be a complex manifold with complex structure $J=B$.
For a tensor $J^i_j$ to be a complex structure, it must satisfy
the conditions $J^2=-1$ and $0=\nabla_{[i} J^{j]}_k$.  The first is
manifest, and given the expression for $J$ in terms of the vector
potential
$$
J^i_j=g^{ik} \partial_{[k} A_{j]}, 
$$
so is the second.

Now, a standard trick to simplify the equations \eq{Dbar-eqns},
is to do a ``gauge transformation'' with a {\it complex} parameter
$\theta(x)$.  While at first this might seem to violate physical
requirements such as the unitarity of the Hamiltonian, in fact it
is perfectly sensible as long as we generalize another ingredient
in the standard definitions, namely the inner product on wave functions.
Explicitly, we define the wave function in terms of another function
$s(x)$, as
\begin{equation}\label{eq:wavef}
\psi(x) = e^{ik\theta(x)} s(x) , \qquad
D_a \psi(x) =
 e^{ik\theta(x)} ~ (i\partial_a + k A_a -k\partial_a\theta) s(x).
\end{equation}
This would be a standard $U(1)$ gauge transformation if 
$\theta(x)$ were real.
By allowing complex $\theta(x)$, and assuming 
$$
0 = [{\bar D}_{\ba},{\bar D}_{\bb}] \equiv F_{\ba\bb}
$$
({\it i.e.} $F^{0,2}=0$), we can find a transformation
which trivializes all the antiholomorphic derivatives,
\begin{equation}
\label{delbar}
{\bar D}_{\ba} \rightarrow \bar\partial_{\ba}.
\end{equation}
In this ``gauge,'' wave functions in the LLL can be
expressed locally in terms of holomorphic
functions.  The only price we pay is that the usual inner product,
$$
\vev{\psi|\psi'} = \int_M \sqrt{g} \psi^*(x) \psi'(x) ,
$$
turns into an inner product which depends on an auxiliary real function,
\begin{equation}\label{eq:h-metric}
h(x) \equiv e^{-2\Im\theta(x)} ,
\end{equation}
as
$$
(s,s') = \int_M \sqrt{g}~ h^k(x) ~ {\bar s}(\bar x) ~s'(x) .
$$
Taking into account the gauge transformations between
coordinate patches, the $s(x)$ are holomorphic sections of a holomorphic
line bundle $L^k$.

In mathematics, one would say that $s(x)$ is a section of $L^k$
evaluated in a specific frame, while the quantity $h^k(x)$ defines a
hermitian metric on the line bundle $L^k$.

\subsection{The lowest Landau level}

Since we have made a one-to-one correspondence between LLL wave
functions and holomorphic sections of the line bundle, we can now find
the total number of LLL states,
which we denote $N$. The number of holomorphic sections
${\rm dim}\, H^0(L^k)$ can be determined for large $k$ by the index
formula \cite{AlvarezGaume:1983at, Friedan:1983xr}
\begin{equation}
\label{eq:dim-H}
N={\rm dim}\,\,H^0(L^k)=\int_M e^{F}\wedge {\rm Td}(M)=a_0k^n+a_1k^{n-1}+\ldots
\end{equation}
where the coefficients $a_i$ are given by certain integrals
involving the curvature of the metric.

Now, given that there is a large degeneracy of ground states and
thus a nontrivial LLL, it becomes interesting to study
the projector on the LLL, or in other words the LLL density matrix
$$
P \equiv \sum_{i; E_i=E_0} \ket{i}\bra{i} .
$$
If we shift $H$ to set the ground state energy $E_0=0$,
it can also be defined as the large time limit of propagation in
Euclidean time. Regarded as a function on two variables, the projector $P$ becomes the Bergman kernel
$$
P(x,x') = \lim_{T\rightarrow\infty} \bra{x}e^{-TH}\ket{x'} .
$$
Thus it can be defined as a path integral by the standard Feynman-Kac formula.

The standard example in which the projector on LLL appears in physics
is the Quantum Hall Effect, see for a review \cite{girvin}. In the
simplest case, one studies 
the dynamics of electrons on a two-dimensional plane
with a constant orthogonal magnetic field. At low
temperatures and high values of the field only the lowest energy level
is important. It is also interesting to consider a partly filled
ground state, with number of fermionic particles $K<N$. In this case
one has to introduce a potential $V$, then particles form an
incompressible droplet, whose edge dynamics is of particular interest.

In recent years this problem has been much generalized;
to Riemann surfaces in \cite{Iengo:1993cs} and
references therein, while
higher dimensional examples include the case of $S^4$
\cite{Zhang:2001xs}, $\mathbb{R}^4$ \cite{Elvang:2002jh} and
$\mathbb{CP}^n$ \cite{Karabali}; see also \cite{Karabali:2006eg} for
a review.

The case of $\mathbb{CP}^n$ is the first nontrivial case in which we can
make contact with the results of this paper. The choice made in
\cite{Karabali} for the $U(1)$ background field is 
\begin{equation}
\label{eq:FS}
F_{a\ba}\sim R_{a\ba},
\end{equation}
proportional to the Ricci tensor. Since for $\mathbb{CP}^n$ the Ricci tensor is equal to the K\"ahler metric, one immediately recognizes \eq{FS} as the
physical condition on the magnetic field \eq{CS}.  Using the local
complex coordinates $z_1,\ldots, z_n$, the LLL wave functions can be
constructed explicitly as
\begin{equation}
\label{wave}
\psi_\alpha\sim\frac{z_1^{\alpha_1}z_2^{\alpha_2}\cdots z_n^{\alpha_n}}{(1+|z|^2)^{k/2}},\quad \alpha=1,\ldots,N
\end{equation}
up to a normalization constant \cite{Karabali:2006eg}. As in \eq{wavef} it has the form of the holomorphic function, weighted by a metric of the line bundle \eq{h-metric}, or, equivalently, the magnetic potential.

The dynamics of the droplet is characterized in the following way. One starts with diagonal density matrix $\rho_0$ with $K$ states occupied, then the fluctuations, preserving the number of states are given by unitary transfomation $\rho_0\rightarrow\rho=U\rho_0U^\dagger$, and the equation of motion is the quantum Liouville equation
$$
i\frac{\p\rho}{\p t}=[V,\rho].
$$
The form of the droplet is determined by the form of the minima of the confining potential. In \cite{Karabali} the case of spherically symmetric potential $V=V(r=z\bz)$ was studied. In the limit of large number of states $N$ (i.e. large magnetic field) and large number of fermions $K<N$ the density matrix has the form
$$
\rho(r^2)=\Theta(r^2-R_d^2),
$$
where $R_d$ is the radius of the droplet and $\Theta$ is the step function. In other words the density matrix is equal to constant in the region, occupied by the droplet. 
This is due to the fact, that the LLL is only partly filled, otherwise it would have been constant everywhere in space. The condition of the constant density matrix (Bergman kernel) turns out to have interesting consequences.

The edge dynamics of the droplet is described by Chern-Simons type action in higher dimensions \cite{Karabali}.

One can generalize the above construction to nonabelian background gauge fields.
Since $\IC\IP^n=SU(n+1)/U(n)$ and Lie algebra of $U(n)=U(1)\times SU(n)$, then in addition to $U(1)$ gauge field one can also turn on $SU(n)$
gauge field. In \cite{Karabali} the case of constant  $SU(n)$
gauge field was considered, so that the field strength is proportional to the $SU(n)$
component of the Riemann curvature. The wave functions (\ref{wave}) as well as the density matrix now carry additional indices, corresponding to $SU(n)$ representation.

The similar generalization of the Bergman kernel was considered recently in \cite{Ma}. 
In addition to the line bundle $L$ one can consider more general hermitian vector bundle $E$ with corresponding connection with the curvature $R^E$. Then the corresponding Bergman kernel is given by the large time limit of the exponential of Dirac operator $D$ squared, for which the expansion analogous to (\ref{exp1}) exists
$$
\rho(x)=\lim_{T\rightarrow\infty}e^{-TD^2}(x,x)=k^n+k^{n-1}(\frac12R\cdot{\bf 1}_E+iR^E)+\ldots.
$$
The second term was computed in \cite{Ma}. It would be
interesting to make further contact between these results and the
higher dimensional Quantum Hall Effect.

\subsection{Applications in K\"ahler geometry}

The original mathematical motivation for this development, usually
attributed to Tian and to Yau, seems to have been to use Bergman
metrics to study the problem of approximation of K\"ahler-Einstein
metrics, which by definition have Ricci tensor proportional to the
metric itself, on complex manifolds.

In \cite{Tian:metrics} Tian considered an algebraic manifold $M$ of
complex dimension $n$, embedded in some projective space
$\IC\IP^N,\,\,N>n$.  Turning on the magnetic field is equivalent to
considering a bundle $L$, or it's $k$-th power $L^k$ for magnetic flux
proportional to $k$, whose choice corresponds to a choice of
``polarization'' on $M$. In local complex coordinates
$z^a,\,\bz^{\ba},\,\,a,\ba=1,\ldots n$ the K\"ahler form $\omega_g$ of
the metric $g_{a\ba}$ is defined as $\omega_g=ig_{a\ba}dz^a\wedge
d\bz^{\ba}$. The K\"ahler metric, polarized with respect to $L$, has
the associated K\"ahler form $\omega_g$ in the same class as the Chern
class $C_1(L)$ of $L$. A particularly useful choice of $\omega_g$ is
to take it to be equal to the curvature of the line bundle (magnetic
field strength). If the hermitian metric of $L$ is $h(z,\bz)$ then for
$L^k$ the metric is $h^k$ and it's curvature is
\begin{equation}
\label{eq:curvature}
kg_{a\ba}=F_{a\ba}=-\p_a\bpartial_{\ba} \log h^k,
\end{equation}
exactly as in \eq{CS}.

Consider next some orthonormal basis $s_0(z),\ldots,s_{N_k}(z)$ on the space $H^0(M,L^k)$ of all global holomorphic sections of $L^k$
\begin{equation}
\label{eq:norm}
(s_\alpha,s_\beta)=\int_M\sqrt gh^ks_\alpha\bs_{\beta}=\delta_{\alpha\beta}.
\end{equation}
One can think of $s_\alpha$ as of projective coordinates on $\IC\IP^{N_k}$. Therefore a particular choice of the basis of sections defines a particular embedding of the manifold $M$ into $\IC\IP^{N_k}$ (different choices of the basis are related by $PGL(N_k+1)$ transformation). The standard metric on the projective space is the Fubini-Study metric $g_{FS}=\p\bpartial \log \sum_\alpha|s_\alpha|^2$. One immediately realizes that $\frac1k$-multiple of it's pullback $\frac1kg_{FS}|_M$ to $M$ is in the same K\"ahler class as the original metric $g$ (\ref{curvature}), since
$$
\frac1kg_{FS}|_M=g+\frac1k\p\bpartial\log\left(h^k\sum_{\alpha=0}^{N_k} s_\alpha\bs_{\alpha}\right)
$$
and the expression inside the logarithm is a globally defined function. This metric is called the Bergman metric. In \cite{Tian:metrics} Tian proved, that as $k\rightarrow\infty$, the Bergman metric converges to $g$ (at least in $C^2$ topology). This result opened a possibility of approximating the K\"ahler-Einstein metrics by the Bergman metrics.   

It is interesting to look at the structure of the ``density of states''
function 
$$
\rho_k(z)=h^k\sum_{\alpha=0}^{N_k} s_\alpha(z)\bs_{\alpha}(\bz).
$$ 
Zelditch \cite{Zelditch:Szego} and Catlin \cite{Catlin} showed that
there is an asymptotic expansion of the density function in $1/k$ in
terms of local invariants of the metric $g$, such as the Riemann
tensor and its contractions.  These invariants were computed by Lu
\cite{Lu} up to third nontrivial order in $1/k$ with the following
result up to the second order in $1/k$
\begin{equation}
\label{eq:Lu}
\rho_k(z)=
 k^n+\frac12k^{n-1}R+k^{n-2}\left(\frac13\Delta
 R+\frac1{24}|\mbox{Riem}|^2-\frac16|\mbox{Ric}|^2+\frac18R^2\right)
 +\CO(k^{n-3})
\end{equation}
The computation is based on the global peak section method, developed
by Tian \cite{Tian:metrics}, which is a
technique to approximate sections of line bundle for large values
of $k$. Another methods to derive this results are the heat kernel approach of \cite{Ma} reproducing kernel approach of \cite{BBS}. Let us also mention their importance for the proof of holomorphic Morse inequalities \cite{Morse,Ma}. 

In this paper, we reproduce the expansion (\ref{eq:Lu}) by taking the
large time limit of the quantum mechanical path integral for a
particle in magnetic field. The function $\rho_k$ is nothing but the
diagonal of the density matrix on the lowest Landau level.

Based on the results of \cite{Tian:metrics,Zelditch:Szego,Catlin,Lu}
Donaldson suggested to
study the metrics with constant density function
$$\rho_k(z)={\rm const}=\frac{{\rm dim}\,H^0(M, L^k)}{{\rm Vol}\,M}.$$ 
Solving the previous equation for $h^k$
and plugging back to the orthonormality condition \eq{norm} we get the equation
$$
\frac{{\rm dim}\,H^0(M, L^k)}{{\rm Vol}\,M}\int_M\sqrt g\frac{s_\alpha\bs_{\beta}}{\sum_{\gamma}s_\gamma\bs_{\gamma}}=\delta_{\alpha\beta}.
$$
on the sections of the line bundle. This is the orthonormality
condition for the basis in $H^0(M,L^k)$. The embedding
$M\rightarrow\IC\IP^{N_k}$, satisfying this condition, is called
``balanced'' \cite{Luo} and the corresponding K\"ahler metric $g_{a\ba}$
\label{curvature} is the ``balanced metric'' (see \cite{Yau} for
the first appearance of this concept). Using the expansion from \eq{Lu} Donaldson was able to show \cite{Donaldson1,Donaldson2,Donaldson:numeric} that under assumption of existence of constant scalar curvature metric, the metric, satisfying previous equation, approaches the metric of constant scalar curvature as $k\rightarrow\infty$. 
In \cite{Donaldson:numeric} an
iterative procedure was proposed to construct these metrics
numerically. One starts with an arbitrary choice of 
basis, parameterized by a hermitian matrix $G_{\alpha\beta}$, and
defines the following integral operator
$$
T(G)_{\alpha\beta}=\frac{{\rm dim}\,H^0(M, L^k)}{{\rm Vol}\,M}\int_M\sqrt g\frac{s_\alpha\bs_{\beta}}{(G^{-1})^{\gamma\delta}s_\gamma\bs_{\delta}}.
$$
The fixed point of this operator $T(G)=G$ corresponds to the balanced
embedding. It was shown in \cite{Donaldson1,Donaldson:numeric} that
for any initial choice of the matrix $G$, the iterative procedure for
$T$ converges to the balanced embedding.  This construction was
recently used for approximating the Ricci flat metrics on Calabi-Yau
hypersurfaces in projective spaces \cite{Douglas:2006rr} and finding
numerical solutions to the hermitian Yang-Mills equation on
holomorphic vector bundles \cite{Douglas:2006hz}.

\section{Non-supersymmetric Bergman Kernel}

\subsection{Density matrix}

The euclidean path integral for a particle on a
$2n$-dimensional K\"ahler manifold $M$ with the magnetic field is
\begin{equation}\label{rho}
\rho(x_i,x_f)=\CN\int^{x(t_f)=x_f}_{x(t_i)=x_i} \prod_{t_i<t<t_f} \det g_{a\bb}(x(t))\, \CD x^a \CD \bx^{\bb}\,\, e^{-\frac1{\hbar}\int_{t_i}^{t_f}dt\,[g_{a\bb}\dot x^a{\dot{\bx}}^{\bb}+A_a\dot x^a+\bA_{\ba}\dot{\bx}^{\ba}]}
\end{equation}
Here we assume that $F_{ab}=F_{\ba\bb}=0$ and work in the anti-holomorphic gauge $A_a=0$ for the gauge
connection\footnote{Although the gauge, which trivializes
anti-holomorphic derivatives is rather $\bA_{\ba}=0$ (\ref{delbar}),
the difference between gauge choices is inessential, since the density
matrix is a gauge invariant object.  We find it convenient to work in
the anti-holomorphic gauge.}. We also set the magnetic field strength to be aligned with the metric
\begin{equation}
\label{eq:balanced}
F_{a\bb}=\partial_a {\bA}_{\bb} =kg_{a\bb}=k\p_a\p_{\bb}K,
\end{equation}
as in \eq{CS}, with $K=-\log h$ being the K\"ahler potential
for the metric. The physical reason for this choice of the magnetic
field strength was outlined in the introduction (namely, to get a
highly degenerate ground state).

Our goal is to compute the value of the density matrix (\ref{rho}) on
the diagonal $x_i=x_f=x$ and in the lowest Landau level, {\it i.e.} in
the large time $T=t_f-t_i\rightarrow\infty$ limit.  However, it turns
out that one cannot take this limit before doing the computation.
If one does this, then the kinetic term in the action
is suppressed, and one obtains the result $1$ for the functional
integral, as is easy to check in first order in $\hbar$.  
Thus, we must keep $T$ finite in the process of calculation and take
the $T\rightarrow\infty$ limit after doing the Feynman integrals. It turns out that this limit is free of IR divergent terms for the choice of magnetic field \eq{balanced} if the path integral is properly regularized. Whether this limit is well-defined for more general field strength, than \eq{balanced}, is an interesting question we will address elsewhere.  

The result is an asymptotic expansion in $\hbar\sim 1/k$, whose
coefficients at each order can be computed using perturbation theory,
and depend on local invariants of the metric, such as the Riemann and
Ricci tensors, curvature scalar and their derivatives.

Let us begin.  To keep track of the $T$ dependence we rescale the
time parameter, defining
$$
t=t_f+(t_f-t_i)\tau=t_f+T\tau
$$ 
with $\tau\in[-1,0]$.
The classical solution for the trajectory 
with boundary conditions $x_i=x_f$ is just a constant function. 
Introduce normal coordinates $z^a,\,\bz^{\ba}$ in the vicinity of the classical trajectory
$$
x^a=x_f^a+z^a(\tau)
$$
$$
\bx^{\ba}=\bx_f^{\ba}+\bz^{\ba}(\tau)
$$
The normalization factor $\CN$ can be fixed by considering the standard normalization of the heat kernel in the case of non-coincident initial and final points, as e.g. in \cite{Bastianelli:1998jb}, and is equal to
$$
\CN=k^n,
$$
where $n$ is the complex dimension of the manifold.

\subsection{Weyl-ordering counterterm}

The path integral representation (\ref{rho}) of the heat kernel 
has been studied 
since the pioneering work by DeWitt \cite{DeWitt:1957at}. 
In the hamiltonian framework the path integral corresponds to 
transition amplitude 
$$
K(x_i,x_f;T)=\la x_f|e^{-(T/\hbar)\hat H}|x_i\ra.
$$
Since the kinetic term in $\hat H$ depends on the coordinate variable through the metric, the well known subtlety arises in this case with the operator ordering of
momentum and coordinate variables 
-- different choices of ordering lead to different lagrangians.
This issue was studied in great detail in
\cite{Bastianelli:1991be,Bastianelli:1992ct,Bastianelli:2006rx}.
There it was shown that there is a convenient choice of the ordering in the hamiltonian which preserves general coordinate invariance
\begin{eqnarray}
\label{hamiltonian}
\nonumber
\hat H &=& \frac12G^{-1/4}(\hat p_i-A_i)G^{ij}G^{1/2}(\hat p_j-A_j)G^{-1/4}\\&&=
\frac12g^{-1/2}\hat p_a\,g^{a\bb}\,g\,(\hat{\bar p}_{\bb}-\bA_{\bb})g^{-1/2}+
\frac12g^{-1/2}(\hat{\bar p}_{\bb}-\bA_{\bb})\,g^{a\bb}\,g\,\hat p_ag^{-1/2},
\end{eqnarray}
where we specified our hamiltonian to the K\"ahler case. Here $G=\det
g_{ij}=g^2= (\det g_{a\ba})^2$. To transform the hamiltonian framework
to lagrangian we rewrite this expression in a Weyl-ordered form (see
Appendix B) and then perform the Legendre transform with the
generalized momenta
$$
p_a=g_{a\bb}{\dot{\bz}}^{\bb},\,\,\,\,\,\,\,\,\,\bp_{\bb}=g_{a\bb}\dot z^a+
\bA_{\bb}.
$$
The following action, written in euclidean time, appears then in the exponent of path integral
$$
S=\int_{t_i}^{t_f}\,dt\,\left(g_{a\bb}\dot z^a{\dot{\bz}}^{\bb}+\bA_{\bb}{\dot{\bz}}^{\bb}-\frac{\hbar^2}{4}R\right).
$$ 
The Weyl ordering corresponds to a ``mid-point rule'' prescription for
path integral representation, which will be introduced in the next
subsection.  The last term in this ``quantum corrected'' action is
necessary e.g. to obtain correct path integral representation for the
small-$T$ heat kernel \cite{Bastianelli:2006rx}. In the next section
we will see that it is also necessary for obtaining the correct
infinite-$T$ expansion.

\subsection{Normal coordinates, free action and propagators}

In K\"ahler normal coordinate frame (see Appendix A for conventions) all
pure (anti-) holomorphic derivatives of the metric at a chosen point
are set to zero. Setting $x=x_i=x_f$, we use K\"ahler normal
coordinates, centered at $x$ and obtain the following expansions for
the K\"ahler potential, metric and gauge connection, up to the sixth
order in derivatives
$$
K(x^a+z^a,\bx^{\ba}+\bz^{\ba})=g_{a\bb}(x)z^a\bz^{\bb}+\frac14K_{ab\ba\bb}(x)z^az^b\bz^{\ba}\bz^{\bb}+\frac1{36}K_{abc\ba\bb\bc}(x)z^az^bz^c\bz^{\ba}\bz^{\bb}\bz^{\bc}+\ldots,
$$
$$
\bA_{\bb}(x^a+z^a,\bx^{\ba}+\bz^{\ba})=k\partial_{\bb}K(x^a+z^a,\bx^{\ba}+\bz^{\ba})
$$
$$
=k\left(g_{a\bb}(x)z^a+\frac12K_{ab\ba\bb}(x)z^az^b\bz^{\ba}+\frac1{12}K_{abc\ba\bb\bc}(x)z^az^bz^c\bz^{\ba}\bz^{\bc}+\ldots\right),
$$
$$
g_{a\bb}(x^a+z^a,\bx^{\ba}+\bz^{\ba})=\partial_a\partial_{\bb}K(x^a+z^a,\bx^{\ba}+\bz^{\ba})=g_{a\bb}(x)+K_{ab\ba\bb}(x)z^b\bz^{\ba}+\frac14K_{abc\ba\bb\bc}(x)z^bz^c\bz^{\ba}\bz^{\bc}+\ldots
$$
in self-explanatory notations. Note that we omitted terms
which turn out not to be relevant up to the second order in
$\hbar$. For example, the term with five derivatives of mixed type
$K_{abc\ba\bb}(x)z^az^bz^c\bz^{\ba}\bz^{\bb}$ is non-zero in our
coordinate frame, but it contributes to the density matrix only
starting from $\hbar^3$, as one can check by power counting.

Using auxiliary ghost fields $b^a$ and $c^{\bb}$ to raise the
determinant from the measure (\ref{rho}) to the exponent, we can 
rewrite the diagonal of the density matrix (\ref{rho}) as
\begin{eqnarray}
\label{density}
\nonumber
\rho(x)&=&\CN\int_{z(-1)=0}^{z(0)=0}\CD z^a(\tau) \CD \bz^{\bb}(\tau) \CD b^a(\tau)\CD c^{\bb}(\tau)\,\,e^{-\frac1{\hbar}S_0-\frac1{\hbar}S_{int}}
\end{eqnarray}
where we split the action into free part
\begin{equation}
\label{free}
S_0=\int_{-1}^0d\tau\,\left[\frac1Tg_{a\bb}(x)\dot z^a{\dot{\bz}}^{\bb}+kg_{a\bb}(x)z^a\dot{\bz}^{\bb}+
g_{a\bb}(x)b^ac^{\bb}\right],
\end{equation}
and interaction part, which up to the sixth order in derivatives of the K\"ahler potential, looks like
\begin{eqnarray}
\label{int}
\nonumber
S_{int}&=&\int_{-1}^0d\tau\,\left[\frac1T\left(K_{ab\ba\bb}(x)z^b\bz^{\ba}+\frac14K_{abc\ba\bb\bc}(x)z^bz^c\bz^{\ba}\bz^{\bc}\right)\dot z^a{\dot{\bz}}^{\bb}\right.
\\&&\nonumber+
\left.
k\left(\frac12K_{ab\ba\bb}(x)z^az^b\bz^{\ba}+\frac1{12}K_{abc\ba\bb\bc}(x)z^az^bz^c\bz^{\ba}\bz^{\bc}\right){\dot{\bz}}^{\bb}\right.
\\&&\nonumber\left.
+\left(K_{ab\ba\bb}(x)z^b\bz^{\ba}+\frac14K_{abc\ba\bb\bc}(x)z^bz^c\bz^{\ba}\bz^{\bc}\right)b^ac^{\bb}\right.
\\&&\nonumber-\left.
\frac{\hbar^2}{4}T\left(R(x)+\partial_{c}\bar\partial_{\bc}R(x)z^c\bz^{\bc}\right)\right],
\end{eqnarray}
and dots denote $\tau$ derivatives. The propagator for free theory (\ref{free}) 
\begin{equation}
\label{zprop}
\la\bz^{\bb}(\tau)z^a(\sigma)\ra=\hbar g^{a\bb}\Delta(\tau,\sigma),
\end{equation}
satisfies equation
$$
\left[-\frac1T\frac{d^2}{d\tau^2}+k\frac{d}{d\tau}\right]\Delta(\tau,\sigma)=\delta(\tau-\sigma),
$$
and the path integral boundary conditions translate into boundary conditions for $\Delta$
$$
\Delta(-1,\sigma)=\Delta(0,\sigma)=\Delta(\tau,-1)=\Delta(\tau,0)=0
$$
The unique solution is 
\begin{eqnarray}
\label{delta}
\nonumber
\Delta(\tau,\sigma)=\frac1{k(e^{kT}-1)}&&\left\{\theta(\tau-\sigma)e^{kT}(1-e^{kT\tau})(1-e^{-kT(\sigma+1)})\right.
\\&&\left.+\theta(\sigma-\tau)(1-e^{-kT\sigma})(1-e^{kT(\tau+1)})\right\},
\end{eqnarray}
where the step-function is defined using the ``mid-point rule''
\begin{eqnarray}
\label{theta}
\theta(\tau-\sigma)=\left\{
\begin{array}{l}
1,\,\,\,\,\tau>\sigma
\\
\frac12,\,\,\,\,\tau=\sigma
\\
0,\,\,\,\,\tau\leq\sigma
\end{array}
\right.
\end{eqnarray}
This value at zero follows from the choice of symmetric ordering in path integral and is crucial for obtaining correct results for heat kernel expansion \cite{Bastianelli:2006rx}.
Ghost propagator can be regulated with the help of $\Delta(\tau,\sigma)$
in the following way
$$
\la b^a(\sigma)c^{\bb}(\tau)\ra=-\hbar g^{a\bb}\delta(\tau-\sigma)=\hbar g^{a\bb}\left(\frac1T\dddel(\tau,\sigma)-k\ddel(\tau,\sigma)\right),
$$
where $\ddel(\tau,\sigma)=d\Delta(\tau,\sigma)/d\tau$, etc.

\subsection{Perturbation theory. First Order}

Now we are ready to study the perturbation theory in $\hbar$ for the
diagonal of the density matrix (\ref{density})
$$
\rho(x)=\CN (1+\hbar\rho_1(x)+\hbar^2\rho_2(x)+\ldots).
$$
From (\ref{zprop}) the dimension of variable $z$ is $\hbar^{1/2}$, therefore by power counting $\rho_n(x)$ should contain 
terms with $2n$ covariant derivatives of the metric. For example, at first order in $\hbar$ the only metric invariant is Ricci scalar.

At the first order in $\hbar$ we have
\begin{eqnarray}
\label{first}\nonumber
\hbar\rho_1&=&-\frac1\hbar K_{ab\ba\bb}\idt\left(\frac1T
\la z^b\bz^{\ba}\dot z^a{\dot{\bz}}^{\bb}\ra|_\tau+
\frac k2\la z^az^b\bz^{\ba}{\dot{\bz}}^{\bb}\ra|_\tau+\la z^b\bz^{\ba}b^ac^{\bb}\ra|_\tau\right)
+\hbar\frac T4R\\&=&\hbar R\frac1T\idt \left(\del(\ddeld+\dddel)+\ddel\deld\right)|_\tau+
\hbar\frac T4R=\hbar RI_1(T,k)+\frac{\hbar T}4R.
\end{eqnarray}
and from here on integration always runs from $-1$ to $0$. Here we
apply usual Wick rule to calculate the correlators and in then we use
the fact that $R_{a\ba b\bb}(x)=K_{ab\ba\bb}(x)$ in normal frame
centered at $x$. This calculation elucidates the role of the ghost
fields $b,\,c$. Namely, their contribution cancels the $\delta(0)$ terms, which
appear in second derivatives of the bosonic propagators at
coinciding points.

The values integrals used in the main text and their large time asymptotics are collected in Appendix C. In $T\rightarrow\infty$ limit of the expression above becomes
\begin{equation}
\label{rho1}
\hbar\rho_1=\left(-\frac T4+\frac1{2k}\right)\hbar R+\frac{\hbar T}4R=\frac\hbar{2k}R.
\end{equation}
Note, that the Weyl-ordering counterterm (\ref{weylhamilt}) is necessary to cancel the large-$T$ divergence. This calculation provides an independent check of the coefficient in front of this term\footnote{Factor $1/4$ here compared to $1/8$ in \cite{Bastianelli:2006rx} is due to our definition of scalar curvature (\ref{Rscalar}) in K\"ahler case.} (see \cite{Bastianelli:2006rx} for detailed consideration of this question).

\subsection{Perturbation theory. Second Order}

At the $\hbar^2$ order the following metric invariants may appear in the expansion:
$\Delta R=g^{a\ba}\p_a\bar{\p}_{\ba}R$, $|\mbox{Ric}|^2=R_{a\ba}R^{a\ba}$, $|\mbox{Riem}|^2=R_{a\ba b\bb}R^{a\ba b\bb}$ and $R^2$.  Therefore the second order correction splits into four components, corresponding to the listed invariants.
The full second-order contribution reads
\begin{eqnarray}
\label{second}
\nonumber
\hbar^2\rho_2=&&-\frac1\hbar K_{abc\ba\bb\bc}\idt\left(
\frac1{4T}\la z^bz^c\bz^{\ba}\bz^{\bc}\dot{z}^a{\dot{\bz}}^{\bb}\ra|_\tau+
\frac k{12}\la z^az^bz^c\bz^{\ba}\bz^{\bc}{\dot{\bz}}^{\bb}\ra|_\tau+
\frac14\la z^bz^c\bz^{\ba}\bz^{\bc}b^ac^{\bb}\ra|_\tau\right)\\
&&\nonumber+\frac1{2\hbar^2}K_{ab\ba\bb}K_{a'b'\ba'\bb'}\idtds\left(
\frac1{T^2}\la z^b\bz^{\ba}\dot{z}^a{\dot{\bz}}^{\bb}|_\tau
z^{b'}\bz^{\ba'}\dot{z}^{a'}{\dot{\bz}}^{\bb'}|_\sigma\ra
\right.\\ &&\nonumber\left.
+\frac kT\la z^b\bz^{\ba}\dot{z}^a{\dot{\bz}}^{\bb}|_\tau
z^{a'}z^{b'}\bz^{\ba'}{\dot{\bz}}^{\bb'}|_\sigma\ra+
\frac{k^2}{4}\la z^az^b\bz^{\ba}{\dot{\bz}}^{\bb}|_\tau
z^{a'}z^{b'}\bz^{\ba'}{\dot{\bz}}^{\bb'}|_\sigma\ra
\right.\\ &&\nonumber\left.
+\frac2T\la z^b\bz^{\ba}\dot{z}^a{\dot{\bz}}^{\bb}|_\tau
z^{b'}\bz^{\ba'}b^{a'}c^{\bb'}|_\sigma\ra+
k\la z^az^b\bz^{\ba}{\dot{\bz}}^{\bb}|_\tau
z^{b'}\bz^{\ba'}b^{a'}c^{\bb'}|_\sigma\ra
\right.\\ &&\left.\nonumber
+\la z^b\bz^{\ba}b^ac^{\bb}|_\tau
z^{b'}\bz^{\ba'}b^{a'}c^{\bb'}|_\sigma\ra
\right)\\&&+\frac{\hbar T}4\partial_c\bar\partial_{\bc}R\idt\la \bz^{\bc}z^c\ra|_\tau+\frac{\hbar T}4R\cdot\hbar
R I_1(T,k)+\frac12\left(\frac{\hbar T}4R\right)^2
\end{eqnarray}
We start computation from the first line in this expression.
Taking into account the identity (\ref{ncidentity}), the first line reads
\begin{eqnarray}
\label{DeltaR}
&&\nonumber-\hbar^2(-\Delta R+2|\mbox{Ric}|^2+|\mbox{Riem}|^2)\idt\frac1T(\ddel\deld\del+\del^2(\ddeld+\dddel)/2)|_\tau
\\&&\nonumber=
-\hbar^2(-\Delta R+2|\mbox{Ric}|^2+|\mbox{Riem}|^2)\,I_2(T,k)
\\&&\approx-\hbar^2(-\Delta R+2|\mbox{Ric}|^2+|\mbox{Riem}|^2)\left(\frac{5}{6k^2}-\frac{T}{4k}\right),\,\,\,{\rm as}\,\,\,T\rightarrow\infty
\end{eqnarray}
Consider now the integral in the second to fifth lines in (\ref{second}). There are several nonequivalent ways to contract $z$ variables, leading to different invariants. Contraction of each of the primed indices $a',\,b',\,\ba',\,\bb'$ with a non-primed index, leads to the $|\mbox{Riem}|^2$ structure. Such terms are given by the following expression
\begin{eqnarray}
\label{riemann}
&& \nonumber\frac{\hbar^2}2|\mbox{Riem}|^2\idtds\left(\frac1{T^2}(
\del(\sigma,\tau)\del(\tau,\sigma)\ddeld(\sigma,\tau)\ddeld(\tau,\sigma)
\right.
\\&&\nonumber\left.+
\del(\sigma,\tau)\deld(\tau,\sigma)\ddeld(\sigma,\tau)\ddel(\tau,\sigma)+
\ddel(\sigma,\tau)\del(\tau,\sigma)\deld(\sigma,\tau)\ddeld(\tau,\sigma)
\right.
\\&&\nonumber\left.+
\ddel(\sigma,\tau)\deld(\tau,\sigma)\deld(\sigma,\tau)\ddel(\tau,\sigma))+
2\frac kT(
\del(\sigma,\tau)\del(\tau,\sigma)\ddeld(\sigma,\tau)\ddel(\tau,\sigma)
\right.
\\&&\nonumber\left.+
\ddel(\sigma,\tau)\del(\tau,\sigma)\deld(\sigma,\tau)\ddel(\tau,\sigma))+
k^2\del(\sigma,\tau)\del(\tau,\sigma)\ddel(\sigma,\tau)\ddel(\tau,\sigma)
\right.
\\&&\nonumber\left.-
\del(\sigma,\tau)\del(\tau,\sigma)\left(\frac1T\dddel(\sigma,\tau)-k\ddel(\sigma,\tau)\right)\left(\frac1T\dddel(\tau,\sigma)-k\ddel(\tau,\sigma)\right)
\right)
\\&&
=\frac{\hbar^2}2|\mbox{Riem}|^2\cdot I_4(T,k)\approx
\frac{\hbar^2}2|\mbox{Riem}|^2\left(\frac7{4k^2}-\frac T{2k}\right),\,\,\,{\rm as}\,\,\,T\rightarrow\infty
\end{eqnarray}
If we contract only two prime and two nonprime indices between each other we get the structure $|\mbox{Ric}|^2$
\begin{eqnarray}
\label{ricci}
\nonumber
\frac{\hbar^2}2|\mbox{Ric}|^2\idtds&&\left(\frac1{T^2}
(\ddel(\tau)\ddel(\sigma)\ddel(\tau,\sigma)\deld(\sigma,\tau)+
\ddel(\tau)\ddeld(\sigma)\del(\tau,\sigma)\deld(\sigma,\tau)
\right.
\\&&\nonumber\left.+
\ddel(\tau)\del(\sigma)\deld(\tau,\sigma)\ddeld(\sigma,\tau)+
\ddel(\tau)\deld(\sigma)\del(\tau,\sigma)\ddeld(\sigma,\tau)
\right.
\\&&\nonumber\left.+
\deld(\tau)\ddel(\sigma)\ddeld(\tau,\sigma)\del(\sigma,\tau)+
\deld(\tau)\ddeld(\sigma)\ddel(\tau,\sigma)\del(\sigma,\tau)
\right.
\\&&\nonumber\left.+
\deld(\tau)\del(\sigma)\ddeld(\tau,\sigma)\ddel(\sigma,\tau)+
\deld(\tau)\deld(\sigma)\ddel(\tau,\sigma)\ddel(\sigma,\tau)
\right.
\\&&\nonumber\left.+
\ddeld(\tau)\ddel(\sigma)\deld(\tau,\sigma)\del(\sigma,\tau)+
\ddeld(\tau)\ddeld(\sigma)\del(\tau,\sigma)\del(\sigma,\tau)
\right.
\\&&\nonumber\left.+
\ddeld(\tau)\del(\sigma)\deld(\tau,\sigma)\ddel(\sigma,\tau)+
\ddeld(\tau)\deld(\sigma)\del(\tau,\sigma)\ddel(\sigma,\tau)
\right.
\\&&\nonumber\left.+
\del(\tau)\ddel(\sigma)\ddeld(\tau,\sigma)\deld(\sigma,\tau)+
\del(\tau)\ddeld(\sigma)\ddel(\tau,\sigma)\deld(\sigma,\tau)
\right.
\\&&\nonumber\left.+
\del(\tau)\del(\sigma)\ddeld(\tau,\sigma)\ddeld(\sigma,\tau)+
\del(\tau)\deld(\sigma)\ddel(\tau,\sigma)\ddeld(\sigma,\tau))
\right.\\
&&\nonumber\left.+
\frac{2k}T
(\ddel(\tau)\ddel(\sigma)\del(\tau,\sigma)\deld(\sigma,\tau)+
\ddel(\tau)\del(\sigma)\del(\tau,\sigma)\ddeld(\sigma,\tau)
\right.
\\&&\nonumber\left.+
\deld(\tau)\ddel(\sigma)\ddel(\tau,\sigma)\del(\sigma,\tau)+
\deld(\tau)\del(\sigma)\ddel(\tau,\sigma)\ddel(\sigma,\tau)
\right.
\\&&\nonumber\left.+
\ddeld(\tau)\ddel(\sigma)\del(\tau,\sigma)\del(\sigma,\tau)+
\ddeld(\tau)\del(\sigma)\del(\tau,\sigma)\ddel(\sigma,\tau)
\right.
\\&&\nonumber\left.+
\del(\tau)\ddel(\sigma)\ddel(\tau,\sigma)\deld(\sigma,\tau)+
\del(\tau)\del(\sigma)\ddel(\tau,\sigma)\ddeld(\sigma,\tau))
\right.
\\&&\nonumber\left.+
k^2(\ddel(\tau)\ddel(\sigma)\del(\tau,\sigma)\del(\sigma,\tau)+
\ddel(\tau)\del(\sigma)\del(\tau,\sigma)\ddel(\sigma,\tau)
\right.
\\&&\nonumber\left.+
\del(\tau)\ddel(\sigma)\ddel(\tau,\sigma)\del(\sigma,\tau)+
\del(\tau)\del(\sigma)\ddel(\tau,\sigma)\ddel(\sigma,\tau))
\right.
\\
&&\nonumber\left.+
\frac2T(\ddeld(\tau)\del(\tau,\sigma)\del(\sigma,\tau)+
\ddel(\tau)\del(\tau,\sigma)\deld(\sigma,\tau)
\right.
\\&&\nonumber\left.+
\deld(\tau)\ddel(\tau,\sigma)\del(\sigma,\tau)+
\del(\tau)\ddel(\tau,\sigma)\deld(\sigma,\tau))
\left(\frac1T\dddel(\sigma)-k\ddel(\sigma)\right)
\right.
\\&&\nonumber\left.+
2k(\ddel(\tau)\del(\tau,\sigma)\del(\sigma,\tau)+
\del(\tau)\ddel(\tau,\sigma)\del(\sigma,\tau))
\left(\frac1T\dddel(\sigma)-k\ddel(\sigma)\right)
\right.
\\
&&\nonumber\left.+
\del(\tau,\sigma)\del(\sigma,\tau)\left(\frac1T\dddel(\tau)-k\ddel(\tau)\right)
\left(\frac1T\dddel(\sigma)-k\ddel(\sigma)\right)
\right.
\\&&\nonumber\left.-
\del(\sigma)\del(\tau)\left(\frac1T\dddel(\sigma,\tau)-k\ddel(\sigma,\tau)\right)\left(\frac1T\dddel(\tau,\sigma)-k\ddel(\tau,\sigma)\right)
\right)
\end{eqnarray}
\begin{eqnarray}
&&
=\frac{\hbar^2}2|\mbox{Ric}|^2\cdot I_5(T,k)\approx
\frac{\hbar^2}{2}|\mbox{Ric}|^2\left(-\frac Tk+\frac3{k^2}\right),\,\,\,{\rm as}\,\,\,T\rightarrow\infty
\end{eqnarray}
If we contract prime indices as well as nonprime indices only between each other, or in other words we contract separately $z$'s and $\bz$'s at point $\tau$ and $z$'s and $\bz$'s at $\sigma$, we get only disconnected diagrams. The structure of this term is just $(\hbar R I_1)^2$. Adding up this term and last two terms from (\ref{second}) we obtain the first order term (\ref{rho1}) squared with the coefficient one-half
\begin{equation}
\label{R^2}
\frac12(\hbar\rho_1)^2=\frac12\left(\hbar RI_1(T,k)+\frac{\hbar T}4R\right)^2\approx\frac{\hbar^2}{8k^2}R^2,\,\,\,{\rm as}\,\,\,T\rightarrow\infty.
\end{equation} 
This term appears since we compute partition function, not the free energy, and therefore do not subtract disconnected diagrams.

Finally the first term in the last line in (\ref{second}) reads
\begin{equation}
\label{dR}
\frac{\hbar^2 T}4\Delta R\idt\del(\tau,\tau)=\frac{\hbar^2 T}4\Delta RI_3(T,k)\approx\hbar^2\Delta R\left(-\frac1{2k^2}+\frac{T}{4k}\right),\,\,\,{\rm as}\,\,\,T\rightarrow\infty.
\end{equation}

Let us now collect all the terms (\ref{DeltaR}, \ref{riemann}, \ref{ricci}, \ref{dR}) that contribute to $\rho_2$ and compute its $T\rightarrow\infty$ limit 
\begin{eqnarray}
\label{rho2}
\nonumber
\rho_2&=&(I_2(T,k)+TI_3(T,k)/4)\Delta R+(-2I_2(T,k)+I_5(T,k)/2)|\mbox{Ric}|^2
\\&&\nonumber+(-I_2(T,k)+I_4(T,k)/2)|\mbox{Riem}|^2+\frac12(I_1(T,k)+T/4)^2R^2
\\&&\approx\frac1{k^2}\left(\frac13\Delta R+\frac1{24}|\mbox{Riem}|^2-\frac16|\mbox{Ric}|^2+\frac18R^2\right),\,\,\,{\rm as}\,\,\,T\rightarrow\infty
\end{eqnarray}
Now we are ready write down the full expansion of the density matrix (\ref{density})
up to second order in $\hbar$
\begin{equation}
\label{exp}
\rho=k^n\left(1+\frac{\hbar}{2k}R+\frac{\hbar^2}{k^2}\left(\frac13\Delta R+\frac1{24}|\mbox{Riem}|^2-\frac16|\mbox{Ric}|^2+\frac18R^2\right)+\CO((\hbar/k)^3)\right).
\end{equation}
Note that this expansion is in perfect agreement with the expansion of Bergman kernel, obtained in \cite{Lu}.

\section{$\CN=(1,1)$ Supersymmetry}

\subsection{Action, symmetries and propagators}

One can obtain expansions similar to (\ref{exp}) in other quantum mechanical theories. 
Here we consider $(1,1)$-supersymmetric particle on K\"ahler manifold with the magnetic field turned on. The action is
\begin{eqnarray}
S=\int_{t_i}^{t_f}\,dt\,\left(g_{a\bb}\dot z^a{\dot{\bz}}^{\bb}+\bpsi^{\ba}(g_{a\ba}\dot{\psi}^a+\dot{x}^b\p_bg_{a\ba}\psi^a)+\bA_{\bb}{\dot{\bx}}^{\bb}+F_{a\ba}\bpsi^{\ba}\psi^a\right)
\end{eqnarray}
This action is invariant under the following $(1,1)$ supersymmetry transformations
\begin{eqnarray}
&&\nonumber\delta x^a=-\bepsilon\psi^a\\
&&\nonumber\delta \bx^{\ba}=-\epsilon\bpsi^{\ba}\\
&&\nonumber\delta \psi^a=\dot{x}^a\epsilon\\
&&\delta \bpsi^{\ba}=\dot{\bx}^{\ba}\bepsilon
\end{eqnarray}
if the metric is K\"ahler and if $A_a,\,\bA_{\ba}$ is a connection of holomorphic vector bundle
$$
F_{ab}=F_{\ba\bb}=0.
$$
Set the field strength to be proportional to the metric, exactly as before in 
\eq{CS}. Consider now the path integral representation of this theory.
If the boundary conditions for $x$ and $\psi$ fields are the same, no ghosts are needed in the action, because bosonic and fermionic determinants cancel in the measure. Moreover, no Weyl-ordering counterterm is needed in this theory, due to fermions.
Bosonic propagator is the same a before, and fermionic propagator 
$$
\la\bpsi^{\bb}(\tau)\psi^a(\sigma)\ra=\hbar g^{a\bb}\Gamma(\tau,\sigma)
$$
satisfies
$$
\left[\frac{d}{d\sigma}+Tk\right]\Gamma(\tau,\sigma)=-\delta(\tau-\sigma)
$$
We would like to compute the ``index density'', i.e. the supertrace of the density matrix,
without performing the $x$-integral
$$
\rho(x)=\lim_{T\rightarrow\infty}\mbox{Str} (-1)^Fe^{-T\hat H}.
$$ 
The right hand side here depends only on the bosonic ``zero-mode''
$x$, and all fermionic dependence is integrated out. Fermion number
insertion $(-1)^F$ corresponds to periodic boundary conditions for
fermions, in which case the propagator has the form
$$
\Gamma(\tau,\sigma)=\frac1{1-e^{kT}}\left(e^{kT(\tau-\sigma)}\theta(\tau-\sigma)+
e^{kT(\tau-\sigma+1)}\theta(\sigma-\tau)\right).
$$

\subsection{Perturbation theory}

The idea of the calculation is the same as in nonsupersymmetric
case. We use K\"ahler normal coordinates and expand the metric around
the constant configuration $x$.

Free part of the action is given by
\begin{equation}
\label{freeN=1}
S_0=\int_{-1}^0d\tau\,\left[\frac1Tg_{a\bb}(x)\dot z^a{\dot{\bz}}^{\bb}+kg_{a\bb}(x)z^a\dot{\bz}^{\bb}+
g_{a\bb}(x)\bpsi^{\bb}\dot{\psi}^a+Tkg_{a\bb}\bpsi^{\bb}\psi^a\right],
\end{equation}
and the interaction part, up to the sixth order in derivatives of the K\"ahler potential, reads
\begin{eqnarray}
\label{intN=1}
\nonumber
S_{int}&=&\int_{-1}^0d\tau\,\left[\frac1T\left(K_{ab\ba\bb}(x)z^b\bz^{\ba}+\frac14K_{abc\ba\bb\bc}(x)z^bz^c\bz^{\ba}\bz^{\bc}\right)\dot z^a{\dot{\bz}}^{\bb}\right.
\\&&\nonumber+
\left.
k\left(\frac12K_{ab\ba\bb}(x)z^az^b\bz^{\ba}+\frac1{12}K_{abc\ba\bb\bc}(x)z^az^bz^c\bz^{\ba}\bz^{\bc}\right){\dot{\bz}}^{\bb}\right.
\\&&\nonumber\left.
+\left(K_{ab\ba\bb}(x)z^b\bz^{\ba}+\frac14K_{abc\ba\bb\bc}(x)z^bz^c\bz^{\ba}\bz^{\bc}\right)\bpsi^{\bb}(Tk+\p_\tau)\psi^a\right.
\\&&\left.+
\left(K_{ab\ba\bb}(x)\bz^{\ba}+\frac12K_{abc\ba\bb\bc}(x)z^c\bz^{\ba}\bz^{\bc}\right)\dot z^b\bpsi^{\bb}\psi^a\right],
\end{eqnarray}

At the first order in $\hbar$ we get
$$
\rho_1(\CN=1)=R\idt\left(\frac1T(\ddel\deld+\ddeld\del)+k\ddel\del-\del\delta(0)+\deld\gam\right)|_\tau=0,
$$
so $\hbar^1$ term is exactly zero, even for finite $T$.

Computation at the second order in $\hbar$ proceeds in a similar fashion as in previous section. Let us only mention one shortcut. Note, that each contraction of $\bpsi(\sigma)$ and $(Tk+\p_{\tau})\psi(\tau)$ is proportional to delta-function $\delta(\tau,\sigma)$, exactly as contraction of ghosts $b$ and $c$. Therefore the first three lines in the interaction lagrangian (\ref{intN=1}) generate the same terms as bosonic interaction lagrangian (\ref{int}) and only the last line in  (\ref{intN=1})
is a new one. With this observation the calculation simplifies significantly. We only give the final answer here
\begin{eqnarray}
\label{secondN=1}
\nonumber
\rho_2(\CN=1)&=&\nonumber-(I_2(T,k)+I_6(T,k))(-\Delta R+2|\mbox{Ric}|^2+|\mbox{Riem}|^2)\\&&\nonumber+
(I_5(T,k)/2+I_7(T,k)+I_8/2)|\mbox{Ric}|^2 \\&&+(I_4(T,k)/2+I_6(T,k)-I_9(T,k))|\mbox{Riem}|^2
\end{eqnarray}
and refer to Appendix C for the values of the integrals here.
The coefficients in front of $|\mbox{Ric}|^2$ and $|\mbox{Riem}|^2$ turn out to be $T$-independent, as a consequence of supersymmetry and the index theorem, and the answer for the density matrix up to the second order in $\hbar$ is
$$
\rho(x)(\CN=1)=k^n\left(1+\frac{\hbar^2}{24k^2}\left(2\Delta R-|\mbox{Ric}|^2+|\mbox{Riem}|^2\right)+\CO(\hbar^3)\right)
$$
This is consistent with the index theorem. According to the latter the $x$-integral of $\rho(x)(\CN=1)$ is equal to the index of Dirac operator on the K\"ahler manifold $M$ for which the exact answer is
$$
\int_{M} dx  \rho(x)(\CN=1)=\mbox{ind} D_A =\int_M \mbox{ch} F\wedge \hat{A}(M).
$$
If we plug $F=kg_{a\bb}dz^a\wedge d\bz^{\bb}$ and expand the A-roof genus  $\hat{A}$ in powers of curvature tensors then the first two terms
in this expression coincide with first two terms in $\int \rho(x)(\CN=1)$, up to maybe an overall constant.

\section{$N=(2,2)$ Supersymmetry}

\subsection{Action, symmetries and propagators}

The action is
\begin{eqnarray}
\nonumber S=&&\int_{t_i}^{t_f}\,dt\,\left(g_{a\bb}\dot z^a{\dot{\bz}}^{\bb}+\bpsi_+^{\ba}(g_{a\ba}\dot{\psi}_+^a+\dot{x}^b\p_bg_{a\ba}\psi_+^a)\right.\\&&\left.+\bpsi_-^{\ba}(g_{a\ba}\dot{\psi}_-^a+\dot{\bx}^b\p_bg_{a\ba}\psi_-^a)+\bA_{\bb}{\dot{\bx}}^{\bb}+F_{a\ba}(\bpsi_+^{\ba}\psi_+^a+\bpsi_-^{\ba}\psi_-^a)\right).
\end{eqnarray}
The $\CN=(2,2)$ supersymmetry transformations
\begin{eqnarray}
&&\nonumber\delta x^a=-\bepsilon_+\psi_+^a-\bepsilon_-\psi_-^a\\
&&\nonumber\delta \bx^{\ba}=-\epsilon_+\bpsi_+^{\ba}-\epsilon_-\bpsi_-^{\ba}\\
&&\nonumber\delta\psi_+^a=\dot{x}^a\epsilon_++\bepsilon_-\Gamma^a_{bc}\psi_-^b\psi_+^c\\
&&\delta\bpsi_+^{\ba}=\dot{\bx}^{\ba}\bepsilon_++\epsilon_-\Gamma^{\ba}_{\bb\bc}\bpsi_-^{\bb}\bpsi_+^{\bc}\\
&&\nonumber\delta\psi_-^a=\dot{x}^a\epsilon_-+\bepsilon_+\Gamma^a_{bc}\psi_+^b\psi_-^c\\
&&\nonumber\delta\bpsi_-^{\ba}=\dot{\bx}^{\ba}\bepsilon_-+\epsilon_+\Gamma^{\ba}_{\bb\bc}\bpsi_+^{\bb}\bpsi_-^{\bc}
\end{eqnarray}
leave the action invariant if connection $A$ is holomorphic
and the hermitian Yang-Mills equation is obeyed
$$
g^{a\bb}D_aF_{b\bb}=0.
$$
Recall again, that our choice of field strength $F_{a\bb}=kg_{a\bb}$
(\ref{eq:CS}) satisfies this equation.

The object that we would like to compute is the Dolbeault index density, which corresponds to taking the supertrace over one species of fermions, and 
setting the zero modes of the second species of fermions to zero.
To achieve this, we choose the following propagators for the fermions
\begin{eqnarray}
\nonumber
\la\bpsi^{\bb}_+(\tau)\psi^a_+(\sigma)\ra&=&\hbar g^{a\bb}\Gamma_+(\tau,\sigma)\\
\nonumber
\la\bpsi^{\bb}_-(\tau)\psi^a_-(\sigma)\ra&=&\hbar g^{a\bb}\Gamma_-(\tau,\sigma),
\end{eqnarray}
where $\Gamma_+$ satisfies periodic boundary conditions: $\Gamma_+(-1,\sigma)=\Gamma_+(0,\sigma)$, $\Gamma_+(\tau,-1)=\Gamma_+(\tau,0)$, and $\Gamma_-$ satisfies Dirichlet b.c. $\Gamma_-(-1,\sigma)=\Gamma_-(\tau,0)=0$. These propagators are given by 
\begin{eqnarray}
\nonumber
\Gamma_+(\tau,\sigma)&=&\frac1{1-e^{kT}}\left(e^{kT(\tau-\sigma)}\theta(\tau-\sigma)+
e^{kT(\tau-\sigma+1)}\theta(\sigma-\tau)\right)\\
\nonumber
\Gamma_-(\tau,\sigma)&=&e^{kT(\tau-\sigma)}\theta(\tau-\sigma),
\end{eqnarray}
One also has to add a pair of bosonic ghost fields $a^a,\ba^{\ba}$, coming from the path integral measure
$$
S_{gh}=\int d\tau g_{a\ba} a^a\ba^{\ba}.
$$

\subsection{Perturbation theory}

The calculation proceeds along the same lines as in the previous two sections. 
Here we present the final answer for the index density 
\begin{eqnarray}
\nonumber
\rho(x)&=&k^n\left(1+\hbar I_{13}(T,k)R+\hbar^2(I_{14}(T,k)\Delta R+(I_5(T,k)/2+I_{11}(T,k)+I_{12}(T,k))|\mbox{Ric}|^2\right.\\\nonumber&+&\left.(-I_2(T,k)+I_4(T,k)/2+I_{10}(T,k))|\mbox{Riem}|^2+I_{13}^2(T,k)R^2/2)+\mathcal O(\hbar^3)\right)\\&=&
k^n\left(1+\frac{\hbar}{2k}R+\hbar^2(I_{14}(T,k)\Delta R-\frac1{6k^2}|\mbox{Ric}|^2+\frac1{24k^2}|\mbox{Riem}|^2+\frac1{8k^2}R^2)+\mathcal O(\hbar^3)\right)\\\nonumber&
\approx&k^n\left(1+\frac{\hbar}{2k}R+\frac{\hbar^2}{k^2}\left(\frac13\Delta R+\frac1{24}|\mbox{Riem}|^2-\frac16|\mbox{Ric}|^2+\frac18R^2\right)+\CO((\hbar/k)^3)\right),\,\,\,{\rm as}\,\,\,T\rightarrow\infty.
\end{eqnarray}

Notice, that the only $T$-dependent term here is a total derivative, as is expected from the index theorem 
$$
\int_{M}\, dx  \rho(x)(\CN=2)=\mbox{ind}\,\, \bar\partial_A =\int_M \mbox{ch} F\wedge \mbox{Td}(M).
$$
Recall, that this index formula computes ${\rm dim}\, \sum_q(-1)^q H^{0,q}(M,L^k)$, which is equal to ${\rm dim}\,H^0(M,L^k)$ for large enough $k$. This explains why $\CN=2$ and nonsupersymmetric Bergman kernel's expansions coincide. 

\section{Conclusions}

In this paper we derived the Tian-Yau-Zelditch {\it et al} expansion
of the Bergman kernel from quantum mechanical path integral. Our
results are in complete agreement with the calculation of
Ref. \cite{Lu}, using Tian's peak section method.

In quantum mechanics, the Bergman kernel corresponds to density matrix
of a particle in strong magnetic field on K\"ahler manifold, projected
to the lowest Landau level. The expansion in the inverse magnetic flux
number can be extracted by taking the infinite time $T\rightarrow\infty$
limit of the (non-supersymmetric) path integral and using the normal
coordinate expansion of the metric and the magnetic field. In this paper we considered a the configuration of magnetic field, being proportional to the K\"ahler form.  Expansions of the same type can also be obtained using supersymmetric quantum
mechanics, where they correspond to index densities. It would be
interesting to extend this result to the case of particle coupled to
non-abelian external fields, considered in mathematical literature
\cite{Ma}.

The physical argument, presented in sec. 2.1, suggests that the analogous expansion for the Bergman kernel may be obtained for a more general magnetic field strength, associated with holomorphic line bundle. We plan to check this in the future publication.

One of the most interesting consequences of this result, as discussed
in section 2, is that there exists a specific magnetic field and
metric for which the density matrix (\ref{exp1}) is constant
everywhere on the manifold.  This is the ``maximally
entropic'' metric for a quantum mechanical observer, in a sense 
discussed in \cite{DK}.

{\bf Acknowledgments} We would like to thank Z. Lu, S. Lukic, S. Lukyanov, B. Shiffman,
G. Torroba, K. van den Broek, P. van Nieuwenhuizen and S. Zelditch for
useful discussions. This work was supported in part by DOE grant
DE-FG02-96ER40959. The work of S.K. was also supported by the grant for support of scientific schools NSh-3035.2008.2 and RFBR grant 07-02-00878. 

\appendix\section{Curvatures}

We follow conventions of \cite{Lu} 
$$
R_{a\ba b\bb}=\partial_b\bar\partial_{\bb}g_{a\ba}-g^{c\bc}\partial_bg_{a\bc}\bar\partial_{\bb}g_{c\ba},
$$
$$
R_{a\ba}=-g^{b\bb}R_{a\ba b\bb},
$$
\begin{equation}
\label{Rscalar}
R=g^{a\ba}R_{a\ba},
\end{equation}
$$
\Delta R= g^{a\ba}\partial_a\bar\partial_{\ba} R,
$$
$$
|\mbox{Riem}|^2=R_{a\ba b\bb}R^{a\ba b\bb},
$$
$$
|\mbox{Ric}|^2=R_{a\ba}R^{a\ba}.
$$

Let $K$ be the K\"ahler potential for the metric
$$
g_{a\ba}=\partial_a\bar\partial_{\ba}K.
$$
In K\"ahler normal coordinate frame the Cristoffel symbols and all pure holomorphic derivatives of the metric vanish at the origin
$$
K_{a\ba b_1\ldots b_n}(x)=0,
$$
for any positive integers $m,\,n$. The following terms in the Taylor expansions of the K\"ahler potential, the metric, Rieman tensor and Ricci scalar are relevant for the present paper
$$
K(x^a+z^a,\bx^{\ba}+\bz^{\ba})=K(x)+K_{a\bb}(x)z^a\bz^{\bb}+\frac14K_{ab\ba\bb}(x)z^az^b\bz^{\ba}\bz^{\bb}+\frac1{36}K_{abc\ba\bb\bc}(x)z^az^bz^c\bz^{\ba}\bz^{\bb}\bz^{\bc}+\ldots,
$$
$$
g_{a\bb}(x^a+z^a,\bx^{\ba}+\bz^{\ba})=g_{a\bb}(x)+K_{ab\ba\bb}(x)z^b\bz^{\ba}+\frac14K_{abc\ba\bb\bc}(x)z^bz^c\bz^{\ba}\bz^{\bc}+\ldots
$$
$$
R_{a\ba b\bb}(x^a+z^a,\bx^{\ba}+\bz^{\ba})=K_{ab\ba\bb}(x)+K_{abc\ba\bb\bc}(x)z^c\bz^{\bc}-g^{c\bc}(x)K_{ab\bc\bd}(x)K_{cd\ba\bb}(x)z^d\bz^{\bd}+\ldots
$$
$$
R(x^a+z^a,\bx^{\ba}+\bz^{\ba})=R(x)+(2g^{a\bd}(x)g^{d\ba}(x)g^{b\bb}(x)K_{ab\ba\bb}(x)K_{cd\bc\bd}(x)
$$
$$
-g^{a\ba}(x)g^{b\bb}(x)K_{abc\ba\bb\bc}(x)-g^{a\ba}(x)g^{b\bb}(x)g^{d\bd}(x)K_{ab\bc\bd}(x)K_{cd\ba\bb}(x))z^c\bz^{\bc}+\ldots
$$
$$
=R(x)+\partial_c\bar\partial_{\bc}R(x)z^c\bz^{\bc}.
$$
The following useful identity holds in the normal coordinate frame
\begin{equation}
\label{ncidentity}
g^{a\ba}g^{b\bb}K_{abc\ba\bb\bc}=-\partial_c\bar\partial_{\bc}R(x)+2R_{c\bd}{R^{\bd}}_{\bc}+R_{c\bb d\bd}{R^{\bb d\bd}}_{\bc}.
\end{equation}

\section{Hamiltonian}

Here we rewrite the hamiltonian (\ref{hamiltonian}) in a Weyl-symmetric way.
First we simplify the expression without the gauge potential
\begin{eqnarray}
\label{weyl}
\hat H &=&
\frac12g^{-1/2}\hat p_a\,g^{a\bb}\,g\,\hat{\bar p}_{\bb}g^{-1/2}+
\frac12g^{-1/2}\hat{\bar p}_{\bb}\,g^{a\bb}\,g\,\hat p_ag^{-1/2}=
\frac12(\hat p_a\,g^{a\bb}\hat{\bar p}_{\bb}+
\hat{\bar p}_{\bb}\,g^{a\bb}\hat p_a)\\&&+\frac{\hbar^2}4\bar\partial_{\bb}(g^{a\bb}\partial_a\ln g)+\frac{\hbar^2}4\partial_a(g^{a\bb}\bar\partial_{\bb}\ln g)+\frac{\hbar^2}4g^{a\bb}\bar\partial_{\bb}\ln g\partial_a\ln g,
\end{eqnarray}
where we use $\hat p_a=-i\hbar \partial_a$, $\hat{\bar p}_{\ba}=-i\hbar\bar\partial_{\ba}$.
The Weyl ordered form of the first term in the previous expression is
$$
(\hat p_ag^{a\bb}\hat{\bar p}_{\bb})_W=\frac14(\hat p_a\hat{\bar p}_{\bb}g^{a\bb}+\hat p_ag^{a\bb}\hat{\bar p}_{\bb}+\hat{\bar p}_{\bb}g^{a\bb}\hat p_a+g^{a\bb}\hat p_a\hat{\bar p}_{\bb})
$$
Therefore
\begin{eqnarray}
\label{weyl1}
\frac12(\hat p_a\,g^{a\bb}\hat{\bar p}_{\bb}+
\hat{\bar p}_{\bb}\,g^{a\bb}\hat p_a)&=&(\hat p_ag^{a\bb}\hat{\bar p}_{\bb})_W+\frac18([\hat p_a,[g^{a\bb},\hat{\bar p}_{\bb}]]+[\hat{\bar p}_{\bb},[g^{a\bb},\hat p_a]])\\&&
=(\hat p_ag^{a\bb}\hat{\bar p}_{\bb})_W+\frac{\hbar^2}4R+\frac{\hbar^2}4g^{a\bb}\Gamma^b_{ab}\Gamma^{\bc}_{\bb\bc}.
\end{eqnarray}
The last three terms in (\ref{weyl}) can be written as
\begin{eqnarray}
\label{weyl2} 
&&
\bar\partial_{\bb}(g^{a\bb}\partial_a\ln g)=\partial_a(g^{a\bb}\bar\partial_{\bb}\ln g)=-R-g^{a\bb}\Gamma^b_{ab}\Gamma^{\bc}_{\bb\bc}.\\&&\nonumber
g^{a\bb}\bar\partial_{\bb}\ln g\partial_a\ln g=
g^{a\bb}\Gamma^b_{ab}\Gamma^{\bc}_{\bb\bc}.
\end{eqnarray}
Using (\ref{weyl1}, \ref{weyl2}) we get the expression for Weyl ordered hamiltonian (\ref{weyl})
\begin{equation}
\label{weylhamilt}
\hat H=(\hat p_ag^{a\bb}\hat{\bar p}_{\bb})_W-\frac{\hbar^2}4R.
\end{equation}
Now it is straightforward to see that the similar expression holds in the presence of gauge connection (\ref{hamiltonian}), one just has to shift $\hat{\bar p}_{\bb}\rightarrow\hat{\bar p}_{\bb}-\bA_{\bb}$ in the previous equation.

\section{Integrals}

Here we collect exact expressions for the integrals that appear in the main text. The following short hand notations are used
\begin{eqnarray}
\nonumber
\ddel(\tau,\sigma)&=&d\del(\tau,\sigma)/d\tau,\,\,\,\,\,\,\,\,
\deld(\tau,\sigma)=d\del(\tau,\sigma)/d\sigma,\,\,\,\,\,\,\,\,
\dddel(\tau,\sigma)=d^2\del(\tau,\sigma)/d\tau,\\\nonumber
\del(\tau,\tau)&=&\del(\tau),\,\,\,\,\,\,\,\,
\ddel(\tau)=\ddel(\tau,\sigma)|_{\sigma=\tau},\,\,\,\,\,\,\,\,
\Gamma(\tau)=\Gamma(\tau,\tau)
\end{eqnarray}
and so on.
\begin{eqnarray}
\nonumber I_1(T,k)&=&\idt\frac1T\left(\del(\ddeld+\dddel)+\ddel\deld\right)|_\tau\\&=&-\frac{e^{kT}+1}{4k(e^{kT}-1)^2}
\left(2+kT+e^{kT}(-2+kT)\right)
\end{eqnarray}
\begin{eqnarray}
I_2(T,k)&=&\idt\frac1T(\ddel\deld\del+\del^2(\ddeld+\dddel)/2)|_\tau=-\frac1{12k^2(e^{kT}-1)^3}\\&&
\nonumber\left(10+3kT+3e^{kT}(6+7kT)+3e^{2kT}(-6+7kT)+e^{3kT}(-10+3kT)\right)
\end{eqnarray}

\begin{eqnarray}
I_3(T,k)&=&\idt\del(\tau,\tau)=\frac1{Tk^2(e^{kT}-1)}\left(2+kT+e^{kT}(-2+kT)\right)
\end{eqnarray}

\begin{eqnarray}\nonumber
I_4(T,k)&=&\idtds\left(\frac1{T^2}(
\del(\sigma,\tau)\del(\tau,\sigma)\ddeld(\sigma,\tau)\ddeld(\tau,\sigma)
\right.\\&&\nonumber\left.+
\del(\sigma,\tau)\deld(\tau,\sigma)\ddeld(\sigma,\tau)\ddel(\tau,\sigma)+
\ddel(\sigma,\tau)\del(\tau,\sigma)\deld(\sigma,\tau)\ddeld(\tau,\sigma)
\right.\\&&\nonumber\left.+
\ddel(\sigma,\tau)\deld(\tau,\sigma)\deld(\sigma,\tau)\ddel(\tau,\sigma))+
2\frac kT(
\del(\sigma,\tau)\del(\tau,\sigma)\ddeld(\sigma,\tau)\ddel(\tau,\sigma)
\right.\\&& \nonumber\left.+
\ddel(\sigma,\tau)\del(\tau,\sigma)\deld(\sigma,\tau)\ddel(\tau,\sigma))+
k^2\del(\sigma,\tau)\del(\tau,\sigma)\ddel(\sigma,\tau)\ddel(\tau,\sigma)
\right.\\&&\nonumber\left.-
\del(\sigma,\tau)\del(\tau,\sigma)\left(\frac1T\dddel(\sigma,\tau)-k\ddel(\sigma,\tau)\right)\left(\frac1T\dddel(\tau,\sigma)-k\ddel(\tau,\sigma)\right)
\right)\\
&=&\nonumber\frac1{4k^2(e^{k T}-1)^4}\left(7 + 2 k T + 12 k Te^{k T}   + 2 e^{2 k T} (-7 + 2 k^2 T^2)\right.\\
&&\left.-12k T e^{3 k T}+ 
 e^{4 k T} (7 - 2 k T)\right)
\end{eqnarray}

\begin{eqnarray} \nonumber
I_5(T,k)&=&\idtds\left(\frac1{T^2}
(\ddel(\tau)\ddel(\sigma)\ddel(\tau,\sigma)\deld(\sigma,\tau)+
\ddel(\tau)\ddeld(\sigma)\del(\tau,\sigma)\deld(\sigma,\tau)
\right.\\&&\nonumber\left.+
\ddel(\tau)\del(\sigma)\deld(\tau,\sigma)\ddeld(\sigma,\tau)+
\ddel(\tau)\deld(\sigma)\del(\tau,\sigma)\ddeld(\sigma,\tau)
\right.\\&&\nonumber\left.+
\deld(\tau)\ddel(\sigma)\ddeld(\tau,\sigma)\del(\sigma,\tau)+
\deld(\tau)\ddeld(\sigma)\ddel(\tau,\sigma)\del(\sigma,\tau)
\right.\\&&\nonumber\left.+
\deld(\tau)\del(\sigma)\ddeld(\tau,\sigma)\ddel(\sigma,\tau)+
\deld(\tau)\deld(\sigma)\ddel(\tau,\sigma)\ddel(\sigma,\tau)
\right.\\&&\nonumber\left.+
\ddeld(\tau)\ddel(\sigma)\deld(\tau,\sigma)\del(\sigma,\tau)+
\ddeld(\tau)\ddeld(\sigma)\del(\tau,\sigma)\del(\sigma,\tau)
\right.\\&& \nonumber\left.+
\ddeld(\tau)\del(\sigma)\deld(\tau,\sigma)\ddel(\sigma,\tau)+
\ddeld(\tau)\deld(\sigma)\del(\tau,\sigma)\ddel(\sigma,\tau)
\right.\\&&\nonumber\left.+
\del(\tau)\ddel(\sigma)\ddeld(\tau,\sigma)\deld(\sigma,\tau)+
\del(\tau)\ddeld(\sigma)\ddel(\tau,\sigma)\deld(\sigma,\tau)
\right.\\&& \nonumber\left.+
\del(\tau)\del(\sigma)\ddeld(\tau,\sigma)\ddeld(\sigma,\tau)+
\del(\tau)\deld(\sigma)\ddel(\tau,\sigma)\ddeld(\sigma,\tau))
\right.\\&&\nonumber\left.+
\frac{2k}T
(\ddel(\tau)\ddel(\sigma)\del(\tau,\sigma)\deld(\sigma,\tau)+
\ddel(\tau)\del(\sigma)\del(\tau,\sigma)\ddeld(\sigma,\tau)
\right.\\&&\nonumber\left.+
\deld(\tau)\ddel(\sigma)\ddel(\tau,\sigma)\del(\sigma,\tau)+
\deld(\tau)\del(\sigma)\ddel(\tau,\sigma)\ddel(\sigma,\tau)
\right.\\&&\nonumber\left.+
\ddeld(\tau)\ddel(\sigma)\del(\tau,\sigma)\del(\sigma,\tau)+
\ddeld(\tau)\del(\sigma)\del(\tau,\sigma)\ddel(\sigma,\tau)
\right.\\&& \nonumber\left.+
\del(\tau)\ddel(\sigma)\ddel(\tau,\sigma)\deld(\sigma,\tau)+
\del(\tau)\del(\sigma)\ddel(\tau,\sigma)\ddeld(\sigma,\tau))
\right.\\&&\nonumber\left.+
k^2(\ddel(\tau)\ddel(\sigma)\del(\tau,\sigma)\del(\sigma,\tau)+
\ddel(\tau)\del(\sigma)\del(\tau,\sigma)\ddel(\sigma,\tau)
\right.\\&&\nonumber\left.+
\del(\tau)\ddel(\sigma)\ddel(\tau,\sigma)\del(\sigma,\tau)+
\del(\tau)\del(\sigma)\ddel(\tau,\sigma)\ddel(\sigma,\tau))
\right.\\&& \nonumber\left.+
\frac2T(\ddeld(\tau)\del(\tau,\sigma)\del(\sigma,\tau)+
\ddel(\tau)\del(\tau,\sigma)\deld(\sigma,\tau)
\right.
\\
&&\nonumber\left.+
\deld(\tau)\ddel(\tau,\sigma)\del(\sigma,\tau)+
\del(\tau)\ddel(\tau,\sigma)\deld(\sigma,\tau)\left(\frac1T\dddel(\sigma)-k\ddel(\sigma)\right)
\right.\\&&\nonumber\left.+
2k(\ddel(\tau)\del(\tau,\sigma)\del(\sigma,\tau)+
\del(\tau)\ddel(\tau,\sigma)\del(\sigma,\tau)\left(\frac1T\dddel(\sigma)-k\ddel(\sigma)\right)
\right.\\&&\nonumber\left.+
\del(\tau,\sigma)\del(\sigma,\tau)\left(\frac1T\dddel(\tau)-k\ddel(\tau)\right)
\left(\frac1T\dddel(\sigma)-k\ddel(\sigma)\right)
\right.\\&&\nonumber\left.-
\del(\sigma)\del(\tau)\left(\frac1T\dddel(\sigma,\tau)-k\ddel(\sigma,\tau)\right)\left(\frac1T\dddel(\tau,\sigma)-k\ddel(\tau,\sigma)\right)
\right)=\\
\nonumber&=&\frac1{k^2(e^{k T}-1)^4}
\left(3 + k T+ e^{k T} (4 + 8 k T + k^2 T^2) + 2 e^{2 k T} (-7 + k^2 T^2) +\right.\\&&
\left.+e^{3 k T} (4 - 8 k T + k^2 T^2)  + e^{4 k T} (3 - k T)\right)
\end{eqnarray}

\begin{equation}
I_6(T,k)=\idt \gam\del\deld|_{\tau}=\frac{1 + e^{k T}}{4k^2(-1 + e^{k T})^3}(3 + k T + 4kTe^{k T}+ e^{2 k T} (-3 + k T))
\end{equation}

\begin{eqnarray}
\nonumber
I_7(T,k)&=&\idtds\left(\gam(\sigma,\sigma)\left((
\ddeld(\tau)\deld(\tau,\sigma)\del(\sigma,\tau)+
\deld(\tau)\ddeld(\tau,\sigma)\del(\sigma,\tau)\right.\right.\\&&\nonumber\left.\left.+
\ddel(\tau)\deld(\tau,\sigma)\deld(\sigma,\tau)+
\del(\tau)\ddeld(\tau,\sigma)\deld(\sigma,\tau))/T\right.\right.\\&& \nonumber\left.\left.+
k(\ddel(\tau)\deld(\tau,\sigma)\del(\sigma,\tau)+
\del(\tau)\ddeld(\tau,\sigma)\del(\sigma,\tau))\right.\right.\\&&\left.\left.\nonumber-\delta(0)\deld(\tau,\sigma)\del(\sigma,\tau)
\right)+\delta(\tau,\sigma)\deld(\sigma)\del(\tau)\gam(\tau,\sigma)\right)\\\nonumber
&=&\frac{1 + e^{k T}}{4 k^2(-1 + e^{k T})^4} (-5 - 2 k T +  e^{k T} (5 - 8 k T - 2 k^2 T^2)\\&& +e^{2 k T} (5 + 8 k T - 2 k^2 T^2)+ e^{3 k T} (-5 + 2 k T))
 \end{eqnarray}

\begin{eqnarray}\nonumber
I_8(T,k)&=&\idtds(\deld(\tau,\sigma)\deld(\sigma,\tau)\gam(\tau)\gam(\sigma)-\deld(\tau)\deld(\sigma)\gam(\tau,\sigma)\gam(\sigma,\tau))\\
&=&\frac1{4 k^2(-1 + e^{k T})^3}(1 + e^{k T} (5 + 4 k T)+ e^{2 k T} (-5 + 4 k T) - e^{3 k T})
\end{eqnarray}

\begin{eqnarray}
\nonumber
I_9(T,k)&=&\idtds\deld(\tau,\sigma)\deld(\sigma,\tau)\gam(\tau,\sigma)\gam(\sigma,\tau)\\
&=&\frac{e^{k T}}{k^2(-1 + e^{k T})^4} (-(-1 + e^{k T})^2 + k^2 T^2 e^{k T})
\end{eqnarray}

\begin{eqnarray}
\nonumber
I_{10}(T,k)&=&\idtds \left( -\frac12\deld(t,s)\deld(s,t)(\Gamma_+(t,s)\Gamma_+(s,t)+\Gamma_-(t,s)\Gamma_-(s,t))\right.\\&+&\left.\nonumber\frac{T^2}2\Gamma_+(t,s)\Gamma_+(s,t)\Gamma_-(t,s)\Gamma_-(s,t)\right)\\&=&\frac{e^{k T} }{2 (-1 + e^{k T})^4 k^2}((-1 + e^{k T})^2 - e^{k T} k^2 T^2)
\end{eqnarray}

\begin{eqnarray}
\nonumber
I_{11}(T,k)&=&\idtds \left( \deld(t,s)\deld(s,t)(\Gamma_+(t)+\Gamma_-(t))(\Gamma_+(s)+\Gamma_-(s))\right.\\&-&\nonumber\left.\deld(t)\deld(s)(\Gamma_+(t,s)\Gamma_+(s,t)+\Gamma_-(t,s)\Gamma_-(s,t))\right.\\&-&\nonumber\left.T^2(\Gamma_+(t)\Gamma_+(s)\Gamma_-(t,s)\Gamma_-(s,t)+\Gamma_-(t)\Gamma_-(s)\Gamma_+(t,s)\Gamma_+(s,t))\right.\\&+&\nonumber\left.\frac2T(\Gamma_+(s)+\Gamma_-(s))
(\deld(t,s)\deld(s,t)\ddel(t)+(\ddeld(t)-T\delta(0))\deld(t,s)\del(s,t)\right.\\&+&\nonumber\left.(\ddeld(t,s)-T\delta(t-s))\deld(s,t)\del(t)+(\ddeld(t,s)-T\delta(t-s))\del(s,t)\deld(t))\right.\\&+&\nonumber\left.2k(\Gamma_+(s)+\Gamma_-(s))
((\ddeld(t,s)-T\delta(t-s))\del(s,t)\del(t)+\deld(t,s)\del(s,t)\ddel(t))\right.\\&+&\nonumber\left.2T\deld(t)(\Gamma_+(t,s)\Gamma_+(s,t)\Gamma_-(s)+\Gamma_-(t,s)\Gamma_-(s,t)\Gamma_+(s))\right)\\&=&\nonumber-\frac1{(-1 + e^{k T})^4 k^2 }(3 (-1 + e^{k T})^2 (1 + e^{k T}) + 
  k T (1 + e^{3 k T} (-3 + k T)\\&&+ e^{k T} (5 + k T) + e^{2 k T} (-3 + 2 k T)))
\end{eqnarray}

\begin{eqnarray}
\nonumber
I_{12}(T,k)&=&\idt\left(\del(\deld+kT\del)(\Gamma_-+\Gamma_+)-\frac1T((\ddel-T\delta(0))\del^2+2\del\deld\ddel)-k\ddel\del^2\right)|_\tau\\&=&\nonumber\frac1{6 (-1 + e^{k T})^3 k^2}(1 + 9 e^{k T} (2 + k T)+ 9 e^{2 k T} (-1 + 2 k T)\\&& + 
 e^{3 k T} (-10 + 3 k T))
\end{eqnarray}

\begin{eqnarray}
\nonumber
I_{13}(T,k)&=&\idt \left(\frac1T((\ddeld-T\delta(0))\del+\ddel\deld)+k\ddel\del-T\Gamma_+\Gamma_-+\deld(\Gamma_++\Gamma_-)\right)|_\tau\\
&=&\frac1{2k}
\end{eqnarray}

\begin{eqnarray}
\nonumber
I_{14}(T,k)&=&\idt \left(\frac1{2T}((\ddeld-T\delta(0))\del^2+2\del\ddel\deld)+\frac k2\ddel\del^2-T\del\Gamma_+\Gamma_-+\del\deld(\Gamma_++\Gamma_-)\right)\\
&=&\frac1{6 (-1 + e^{k T})^3 k^2}(1 - 6 e^{k T} - 3 e^{2 k T} (-1 + 2 k T)+ 2 e^{3 k T} )
\end{eqnarray}

\end{document}